\documentclass[]{article}

\usepackage{amssymb,latexsym,amsmath}     
\usepackage{authblk}
\usepackage[utf8]{inputenc}
\usepackage{dsfont}
\usepackage[utf8]{inputenc}
\usepackage{graphicx}
\usepackage{mathtools}
\usepackage{amsmath}
\usepackage{subfig}
\usepackage{xcolor}
\usepackage{epstopdf}
\usepackage[colorlinks = true,
linkcolor = blue,
urlcolor  = blue,
citecolor = blue,
anchorcolor = blue]{hyperref}
\usepackage[numbers,square,sort&compress]{natbib}
\usepackage{pifont}

\begin{document}
\title{Computational Complexity in Analogue Gravity}

\author{Shahrokh Parvizi \thanks{parvizi@modares.ac.ir} }
\author{Mojtaba Shahbazi \thanks{mojtaba.shahbazi@modares.ac.ir} }

\affil{Department of Physics, School of Sciences,
	Tarbiat Modares University, \\P.O.Box 14155-4838, Tehran, Iran}

\maketitle
\begin{abstract}
Analogue gravity helps to find some gravitational systems which are similar to the evolution of perturbation in condensed matter systems. These analogies provide a  very good tool for either side. In other words, some aspects of gravity could be simulated in condensed matter laboratories. In this study, we find an interpretation for computational complexity in condensed matter systems in terms of the flux density of the fluid and the analogue of the uncertainty principle as the Lloyd bound. We show that the Lloyd bound is reduced to the shear viscosity to entropy ratio (SVER). It has been revealed that the analogue gravity is a fluid located at a time-like finite cut-off surface (call it the bulk fluid) and we found the relation between SVER of the analogue gravity and the boundary fluid. Then we see that whenever the KSS bound is satisfied in the boundary fluid, the KSS bound could be either satisfied in the bulk fluid or not; in addition, when the KSS bound is violated in the boundary fluid, then the KSS bound is violated in the bulk fluid. In other words the satisfaction of the KSS bound in the boundary fluid is a necessary condition for the satisfaction of the KSS bound in the bulk fluid.
\end{abstract}

\section{Introduction}
The AdS/CFT duality allows us to quantitatively study some specific geometric quantities in the gravity side, which correspond to some operators in the quantum field theory side and vice versa \cite{mal}. Recently a new concept has been considered in which the quantum computational complexity corresponds to a geometrical quantity in the bulk. Computational complexity is a very well-known idea in the quantum information theory, which states complexity is the minimum number of quantum gates required to form a target state from a reference state \cite{comp}. The holographic picture of the notion developed by different conjectures such as complexity equal action proposal \cite{ca}, known as the CA proposal, corresponds to the action of a gravitational theory in Wheeler-De Witt patch (WDW):
\begin{align}
C=S_{WDW}\;.
\end{align}
Another conjecture is complexity equal volume \cite{shock}, known as the CV proposal that relates the volume of the Einstein-Rosen bridges (ERB), a geometric notion to the computational complexity in the quantum side.
A modification to the CV proposal is the so-called CV-2.0 proposal that states complexity is equal to thermodynamic pressure multiplied by the volume of the WDW patch \cite{couch}.

In addition, Lloyd \cite{lloyd} introduced an upper limit for the rate of complexity change which is called the Lloyd limit:
\begin{equation}
\frac{dC}{dt} \le 2M \label{lloyd}\;,
\end{equation}
where $M$ is the mass of the black hole. The Lloyd limit is violated in several theories \cite{myers,Moosa,Ghodrati,Yang,alish} for CA and CV conjectures. However, the CV2.0 proposal reaches more successes, particularly in Einstein-Maxwell-Dilaton theories \cite{couch,lc}, and computations in this proposal are analytical rather than numerical, particularly in the complexity growth rate.

It is fascinating to find an interpretation for the notion of complexity in condensed matter systems since it broadens our understanding of the concept of complexity. In addition, it makes it possible to study the systems experimentally. Since some gravitational phenomena are not available to be tested experimentally, simulation of those by some available analogue systems could be useful, particularly when it leads to technology.

Formation of a black hole requires mass heavier than the solar mass as a consequence of Chandrashekhar limit. For a black hole of solar mass (or heavier), the Hawking temperature is about (or less than) $10^{-8}K$ which is much smaller than the cosmic microwave background (CMB) temperature about $2.7 K$ and the lifetime of these black holes scales as the age of the universe. It means that the detection of black holes of solar mass or heavier is impossible by our cutting-edge technology. However, it seems that there is a chance in primordial black holes (PBH) which are formed just after the Big Bang and have lighter masses. Although the evaporation of PBHs could generates gravity waves which is not detectable by virtue of current observatories \cite{gravitywave}, those PBHs being at the final stage of their Hawking radiation release much energy and have bright flashes which could be detected astronomically by X-ray future experiments \cite{xray}, future opportunities from the radiation of PBH dark matter nearby dark matter dense regions \cite{dark}. Nonetheless, the future chances of PBHs observation have nothing to do with the interior of the black holes, whereas, the complexity has the advantage of being a probe to study the interior region. In such manner, analogue black holes are distinct and worth considering by way of simulating the inside of black holes.
 For example, in \cite{hayden} it showed that the interior of a black hole could be equated to a quantum circuit, in other words, a black hole is a natural quantum computer and even it is the fastest quantum computer in nature \cite{scram}. The outcome of this quantum computer is the Hawking radiation. In addition, Weinfurtner et al. have detected the analogue of the Hawking radiation in a water tank experiment \cite{water}, then supersonic black holes in analogue gravity can be equated to an analogue quantum computer. The other applications of analogue gravity in technology are found in \cite{cloak,a,exp}. 
In addition, analogue gravity interpretation could provide a perspective to resolve theoretical physics problem. For example, it has been revealed that information loss in black hole radiation could be interpreted as momentum loss over the analogue black hole horizon and as a result this interpretation provides a non-ad hoc resolution to the information paradox \cite{infoana}. Such interpretation presented in this work, enables us to relate the Lloyd bound in information theory to the KSS bound on the ratio of viscosity to entropy density in the context of fluid-gravity duality and correspondingly relate the information theory to the analogue black holes.

The AdS/CFT duality makes an avenue to relate gravity theories to condensed matter systems. A successful prototypical case is superconductivity in which one applies gravity methods to superconductors \cite{sc,sc1}. The other way to correspond a gravity theory to a condensed matter system is the analogue gravity. It has been seen that excitations over the background in certain condensed matter systems obey equations of motion that are analytically identical to that of perturbation traveling in a gravity theory with a metric. In this manner, one can see the gravitational effects in condensed matter systems and experimentally test them. As an interesting example, the Hawking radiation has been realized by the analogue gravity \cite{a}. 

We therefore can realized two different pictures which relate the analogue fluid to the holographic boundary fluid as follows:
\begin{enumerate}
\item One can consider a conformal fluid, call it $F$ and correspond it to a gravitational theory call it $G_a$ via analogue gravity correspondence. On the other hand, the fluid $F$ can be related by the holography to an AdS geometry $G_h$. Although, $F$ is the same in both cases, $G_a$ and $G_h$ are different geometries, in this way we have two alternative description of the same fluid \cite{conformalana,dumb}.  
\item One can consider a gravitational theory call it $G$ and correspond it to a fluid in analogue gravity $F_a$ in one hand, and correspond it to a boundary fluid $F_h$ via holography, on the other hand. It has been revealed that $F_a$ is at a time-like cut-off surface in the $AdS$ background, and in this vein, $F_a$ and $F_h$ describe two different fluids. Nonetheless, if a black hole in the bulk has thermodynamic properties $\rho, T, ...$, then $F_h$ shares the same thermodynamic properties, but $F_a$ has different ones which are related to that of $F_h$. More on this is discussed in section \ref{realvana}
\end{enumerate}

Our purpose in this paper is following the second way, taking a particular gravitational theory and extend some notions into the analogue counterparts
and find an interpretation for computational complexity in condensed matter systems by analogue gravity. In section 2, the analogue gravity is reviewed, then in section 3, we find the complexity of an acoustic black brane by the CV2.0 proposal and the Lloyd bound is considered. In this section, we also provide some experimental criteria to verify the validity of the Lloyd bound simulation in a laboratory. In section 4, the complexity of the black brane is computed in terms of quantities in condensed matter. We also find a relation between complexity in an acoustic black brane and complexity in a real black brane in the AdS bulk. We show the connection 
between the Lloyd bound and the viscosity to entropy ratio bound.

\section{Analogue Gravity}\label{review}
Some condensed matter systems resemble an analogue for gravity such that small perturbations around the background obey equations of motion identical to that of fields in a curved spacetime. The effective metric in the condensed matter system of analogue gravity is known in terms of the flow and density of the background. In this way, some certain forms of spacetime backgrounds can be interpreted as an analogue in a condensed matter system. Some examples could be simulated with gravity analogs such as expanding de-Sitter space \cite{des}, Schwarzschild black hole \cite{sch} and black brane \cite{bb}. The allowed dimensions are $3+1$ for the conformal Schwarzschild black hole, and $4+1$ for the black brane \cite{bb}. Nonetheless, one can have a larger class of conformal Schwarzschild and black brane spacetimes at the expense of having an effective mass for the perturbation \cite{ac}.

For analogue gravity to work, one requires a condensed matter system that generates the desired metric field; in addition, it has to fulfill its equations of motion.
The analog metric could be written using the Lagrangian approach. The Lagrangian of a fluid $\mathcal{L}$ in $m+1$ dimensions is read as\footnote{A general Lagrangian should contain a vectorial, tensorial and spinorial modes. However, for the sake of simplicity we consider only the scalar mode.}:
\begin{align} 
\mathcal{L}=\mathcal{L}(\eta^{\mu \nu} \partial_\mu \phi \partial_\nu \phi -V(\phi ,t,x))=\mathcal{L}(\mathcal{K} -V(\phi ,t,x))\label{lagrang}\;.
\end{align} 
The energy tensor by variational principle could be written as:
\begin{align} 
T_{\mu \nu}=-\Big(2\frac{\partial \mathcal{L}}{\partial \mathcal{K}}\partial_\mu \phi \partial _\nu \phi-\mathcal{L}\eta_{\mu \nu}\Big)\;,
\end{align} 
and by the energy tensor for fluid dynamics:
\begin{align} 
T_{\mu \nu}=(p+\rho) u_\mu u_\nu +p \eta_{\mu \nu}\label{energyten}\;,
\end{align} 
one can represent the acoustic metric in terms of quantities in fluid dynamics:
\begin{align} 
u_\mu&=\frac{\partial _\mu \phi}{\sqrt{\mathcal{K}}} \\
p&=\mathcal{L} \\ 
\rho&=2\mathcal{K}\frac{\partial\mathcal{L}}{\partial\mathcal{K}}-\mathcal{L}\;.
\end{align} 
If one perturbs the scalar field around a background $\phi=\phi_0+\epsilon \phi_1$, the Lagrangian up to the second order could be written as follows:
\begin{equation}
\mathcal{L}=\mathcal{L}_0+\epsilon \mathcal{L}_1+\epsilon^2\mathcal{L}_2+\mathcal{O}(\epsilon^3)\;.
\end{equation}
By integrating by parts, the term of the second order would be (the first order leads to the equation of motion):
\begin{align}
S=&S_0+\frac{\epsilon^2}{2}\int d^{m+1}x\Big(\big(\frac{\partial^2\mathcal{L}}{\partial(\partial_{\mu}\phi)\partial(\partial_{\nu}\phi)}\big)\partial_{\mu}\phi_1 \partial_{\nu}\phi_1+\big(\frac{\partial^2\mathcal{L}}{\partial\phi \partial\phi}-\partial_{\mu}(\frac{\partial^2\mathcal{L}}{\partial(\partial_{\mu}\phi) \partial\phi})\big)\phi_1\phi_2\Big)+ \mathcal{O}(\epsilon^3)\\
:=&S_0+\frac{\epsilon^2}{2}\int d^{m+1}x \Big(\sqrt{-g}g^{\mu\nu}\partial_{\mu}\phi_1 \partial_{\nu}\phi_1-\sqrt{-g} m_{eff}^2 \phi_1^2\Big)\;.
\end{align}
By use of the chain rule:
\begin{align}
\frac{\partial}{\partial(\partial_{\mu}\phi)}=\frac{\partial \mathcal{K}}{\partial(\partial_{\mu}\phi)}\frac{\partial}{\partial\mathcal{K}}=2\eta^{\mu\nu}\partial_{\nu}\phi \frac{\partial}{\partial\mathcal{K}}\;,
\end{align}
the term $\sqrt{-g}g^{\mu\nu}$ is written as:
\begin{equation}
\sqrt{-g}g^{\mu\nu}=-2\Big(\eta^{\mu\nu}\frac{\partial\mathcal{L}}{\partial \mathcal{K}}-2\mathcal{K}u^{\mu}u^{\nu}\frac{\partial^2\mathcal{L}}{\partial \mathcal{K}^2}\Big)\;. \label{dm}
\end{equation}
Now, we define $c^{-2}=\frac{\partial \rho}{\partial p}$ where $c$ is the speed of sound. Considering the fluid in the rest frame $u_{\mu}=(1,\vec{0})$ and use of eq. \eqref{dm}, the determinant of $g_{\mu \nu}$ is given as follows \cite{bb}:
\begin{equation}
\sqrt{-g}=c^{-\frac{2}{m-1}}\Big(-\frac{\rho+p}{\mathcal{K}}\Big)^{\frac{m+1}{m-1}}\;.
\end{equation}
Then the acoustic metric is read:
\begin{align} 
g_{\mu \nu}=c^{\frac{-2}{m-1}}\Big(\frac{\rho+p}{\mathcal{K}}\Big)^{\frac{2}{m-1}}\Big(\eta_{\mu \nu}+(1-c^2)u_\mu u_\nu\Big) \label{am}\;.
\end{align} 

In the non-relativistic limit where $p\ll \rho$ and $v^2\ll c_l^2$ (where $c_l$ is the speed of light)\footnote{Throughout the paper we set the speed of light $c_l=1$}, the line element would be \cite{ac}:
\begin{equation}
ds^2= \Big(\frac{\rho}{\mathcal{K}c}\Big)^{\frac{2}{m-1}}\Big(-c^2dt^2+\delta_{ij}(dx^i-v^i dt)(dx^j-v^j dt)\Big) \label{aco}\;,
\end{equation}
and the equations of motion for the background are the continuity and the Euler equations:
\begin{align} 
\partial_t \rho+\nabla \cdot(\rho\vec{v})&=0 \label{coneq}  \\
\rho \Big(\partial_t \vec{v}+(\vec{v}\cdot\nabla)\vec{v}\Big)&=\vec{F}\;.
\end{align} 
The kinetic term which is interpreted as specific enthalpy, could be written in terms of the fluid quantities \cite{bb}:
\begin{align} 
\mathcal{K}\sim \Big(\frac{\rho+p}{n}\Big)^{2}\sim (m_0a)^2\;,
\end{align} 
where $m_0$ is the particles mass, $a$ a constant of energy dimension and $n$ the particle density in pressure $p$ . Furthermore, one can show that $\mathcal{K}$ in relativistic limit is given by \cite{particle}:
\begin{align} 
&\mathcal{K}=\frac{1}{\rho_{0} c_l }\Big(\frac{(\rho+p)n_{0}}{n }\Big)^{2}\;,\\ \label{ent}
&n(p)=n_{0}\exp\Big(\int_{\rho_0}^{\rho(p)} \frac{d\rho}{\rho+p}\Big)\;,
\end{align} 
where $n_{0}$ and $\rho_{0}$ are particle density and energy density in zero pressure limit, respectively.

Since we are looking for an interpretation for complexity through AdS/CFT, we are going to see how AdS black holes mimic the acoustic metric. The idea is that there is a correspondence between the quantum theory and the semi-classical gravity through AdS/CFT. Moreover, there is a correspondence between gravity and condensed matter systems, say, a fluid through analogue gravity. We are going to take advantage of a correspondence between the quantum theory and a condensed matter system.

This idea is different from the famous fluid/gravity duality where a strongly coupled fluid is related to a semi-classical gravity. In short, fluid/gravity duality is a correspondence between gravitational bulk theory and a boundary fluid, while analogue gravity relates a gravitational bulk theory to a fluid on a time-like finite cut-off \cite{ra}. In section \ref{realvana} we discuss it further. 

\subsection{AdS black holes}\label{adssec}
An AdS-Schwarzschild black hole in $d+1$ dimensions is given by:
\begin{align}
&ds^2=-f(r)dt^2+\frac{dr^2}{f(r)}+r^2d\sigma^2_{d-1} \label{ads}\;,\\
&f(r)=k+\frac{r^2}{L^2}-\frac{\mu}{r^{d-2}}\;,\\
&\mu=R^{d-2}\Big(\frac{R^2}{L^2}+k\Big)\label{m}\;,
\end{align}
where $R$ is the horizon radius, $L$ AdS radius and $k$ is $-1$, $0$ and $1$ in hyperbolic, flat and spherical horizon geometries, respectively. The thermodynamic quantities of the black hole would be:
\begin{align} 
T&=\frac{f'(R)}{4\pi}\label{tem} \;,\\
S&=\frac{A}{4G}=\frac{V_{d-1}}{4G}R^{d-1}\;, \\
M&=\frac{V_{d-1}(d-1)}{16\pi G}\mu \label{thermo}\;.
\end{align} 
Cosmological constant is:
\begin{align}  
\Lambda&=-\frac{(d-1)(d-2)}{2L^2}\;.
\end{align} 

The metric could be turned into an acoustic metric by coordinate transformation $t=t'+\bar{r}(r)$ where:
\begin{align} 
\bar{r}(r)=\int^r dr~\frac{\sqrt{1-f}}{f}\;,
\end{align} 
which leads \cite{bb}:
\begin{align} 
ds^2=-fdt'^2-2\sqrt{1-f}dt'dr+dr^2+r^2d\sigma^2_{d-1}\label{Painsch}\;.
\end{align} 
Then by comparison to metric (\ref{aco}), we can read off:
\begin{align} 
\rho&=\mathcal{K}c\;, \\
v&=\sqrt{1-f}\;.
\end{align}
The above equations are important because the velocity and the density of the fluid should satisfy the continuity equation \eqref{coneq}. For example, it can be shown that for a $3+1$ dimensional Schwarzschild black hole with $k=1,-1$, there is not any analogous fluid and there is analogue for flat geometry $k=0$ in 4+1 dimensions \cite{bb}. Note that we simulated a $d+1$ dimensional gravitational theory with a $m+1$ dimensional fluid where it could be $m\neq d$. These cases are studied in \cite{bb} as simulating a part of the gravitational theory. To generalize the analogue gravity for arbitrary dimensions, one can add a conformal factor to the metric \eqref{ads}:
\begin{align}
ds^2=\Omega^2\Big(-f(r)dt^2+\frac{dr^2}{f(r)}+r^2d\sigma^2_{d-1}\Big) \label{conformal}\;,
\end{align}
which satisfies equation of motion of a scalar field in the background \eqref{conformal}. However, it could be seen that the conformal factor could play the role of an effective mass for the scalar field. In other words, if a massive scalar field in the background \eqref{conformal} satisfies the Euler equation in fluid dynamics $\mathcal{L}(\phi', g'_{\mu\nu}=\Omega^2 g_{\mu\nu})=\mathcal{L}(\rho, p, v)=0$, then a massive scalar field with a new effective mass in the background \eqref{ads} satisfies the Euler equation in fluid dynamics $\mathcal{L}(\phi, g_{\mu\nu})=\mathcal{L}(\phi', \Omega^2 g_{\mu\nu})=\mathcal{L}(\rho, p, v)=0$:
\begin{align}
&g'^{\mu\nu}\nabla'_{\mu}\nabla'_{\nu}\phi'-M'^2\phi'=g^{\mu\nu}\nabla_{\mu}\nabla_{\nu}\phi-M^2\phi=0\;,\\
&M^2=\Omega^2M'^2+\Omega^{\frac{2-d}{2}}g'^{\mu\nu}\nabla'_{\mu}\nabla'_{\nu}\Omega^{\frac{d-2}{2}}\;,\\
&\phi=\Omega^{\frac{d-2}{2}}\phi' \;.
\end{align}

It means that a particular fluid $(\rho, p, v)$ is capable of simulating two related field contents $(\phi', g')$ and $(\phi,g)$. Due to the fact that in some cases if $(\phi, g)$ is an asymptotic AdS solution, $(\phi',g')$ is not an asymptotic AdS solution, in the following we go on with $(\phi,g)$ rather than $(\phi', g')$.  Then the fluid quantities of \eqref{conformal} are read:
\begin{align}
&\rho=\mathcal{K}c \Omega^{m-1}\label{roo}\;,\\
&v=\sqrt{1-f}\label{vee}\;,
\end{align}
by the fact that these quantities should satisfy continuity equation, we have:
\begin{align}
&\nabla.(\rho v)=0~\rightarrow \partial_r (r^{m-1}\rho v)=0\;,\\
&r^{m-1}\rho v=cte.=A~\rightarrow \Omega^{m-1}=\frac{A}{r^{m-1}\mathcal{K}c \sqrt{1-f}}\label{omeg}\;.
\end{align}   
Then the fluid flux would be:
\begin{equation}
\rho v=\frac{A}{r^{m-1}}\label{flux}\;.
\end{equation}
In the next section, we show how computational complexity is related to the quantities in fluid dynamics.
\section{CV2.0 Proposal and the Lloyd bound}\label{spr}
In this proposal, computational complexity is related to the volume of spacetime in the WDW domain and pressure of the black hole \cite{couch}:
\begin{align}
C=PV_{WDW} \label{com}\;,
\end{align}
where the pressure is related to the cosmological constant:
\begin{align}
P=-\frac{\Lambda}{8\pi G}=\frac{d(d-1)}{16\pi G L^2} \label{pre}\;.
\end{align}
There is a limit known as the Lloyd bound which constraints the complexity rate of change as \cite{lloyd}:
\begin{align}
\frac{dC}{dt}\le2M\;,
\end{align}
and for the CV2.0 proposal, one finds:
\begin{equation}
\dot{C}=PV\le2M \label{l}\;,
\end{equation}
where $V$ represents a thermodynamic volume that is defined for static solutions as an integration over the black hole interior \cite{couch}. For example, the thermodynamic volume of the AdS-Schwarzschild black hole is:
\begin{align}
V=\int_0^{r_h}\sqrt{-g}dV=\frac{4}{3}\pi r_h^3\;.
\end{align}
Spacetime volume for metric (\ref{ads}) would be:
\begin{align}
V_{WDW}=\int_{WDW} \sqrt{-g}dV=\int_{WDW} r^{d-1}dV\;. \label{comcv2}
\end{align}

Now let us consider the relativistic irrotational fluid and the corresponding emergent metric, then comparing \eqref{am} with \eqref{Painsch}, leads to the following fluid quantities\footnote{Here we insert an conformal factor $\Omega$ in \eqref{Painsch}}:
\begin{align}
rr~component:~&\Omega^2=\Big(\frac{\rho+p}{\mathcal{K}c}\Big)^{\frac{2}{m-1}}\Big(1+(1-c^2)\gamma^2 v^2\Big) \label{rhoflu}\;,\\
rt~component:~&\sqrt{1-f}=\frac{(1-c^2)\gamma^2v}{1+(1-c^2)\gamma^2v^2}\label{veflu}\;,
\end{align}
where $\gamma=\frac{1}{\sqrt{1-v^2}}$. By relativistic continuity equation, $\rho$ and $v$ are related as follows:
\begin{align}
\nabla \cdot (\rho\gamma v)=0\;,
\end{align}
which gets into:
\begin{align}
r^{m-1}\rho \gamma v=cte.=A~\rightarrow ~r=\Big(\frac{A}{\rho \gamma v}\Big)^{\frac{1}{m-1}}\;.
\end{align}

Then \eqref{comcv2} is rewritten:

\begin{align}
V_{WDW}=\int_{WDW} \Big(\frac{A}{\rho \gamma v}\Big)^{\frac{d-1}{m-1}} dV\;,
\end{align}
and for $d=3$ spacetime volume and a fluid in $m=3$ spatial dimensions we have:
\begin{align}
V_{WDW}=\int_{WDW} \frac{A}{\rho \gamma v} dV \label{vo}\;,
\end{align}
complexity in terms of fluid quantities and \eqref{vo} reads as:
\begin{equation}
C=\frac{A}{8\pi G L^2} \int_{WDW} \frac{1}{\rho \gamma v}dV\;. \label{co}
\end{equation}
We call it {\it complexity of the fluid}. Now we have an interpretation for complexity in terms of the fluid. The integrand in \eqref{co} is the inverse of the flux density of the fluid, which means that the complexity in condensed matter systems could be interpreted as the inverse of the density of the fluid flux restricted to the WDW patch\footnote{WDW patch is defined as the union of space-like curves connecting the boundary times in holography. Then, if one is interested in the measuring the complexity of the fluid should measure the flux density of the fluid and integrate the inverse flux on the WDW patch of the bulk gravity. The gravitational bulk theory is simulated by a fluid that mimics effectively the gravitational theory. Then the fluid here effectively is an AdS black hole where the horizon, according to \eqref{am}, is located at $1+(1-\frac{1}{c_s^2})\gamma^2 v^2=0$, $c_s$ is the speed of sound, where the velocity of the fluid is equal to $v^2=\frac{1}{\gamma^2 (\frac{1}{c_s^2}-1)}$. In this manner, WDW patch is restricted by four light-like curves, $ds^2=0$ in \eqref{am}, $\frac{dr^2}{dt^2}=\frac{c^2_{l}}{1+(1-c^2_{s})\gamma^2 v^2}$ where $c_l$ is the speed of light. The boundary of the theory where the corners of the WDW patch are attached to is located on $r\rightarrow \infty$ where the metric reduces to the AdS spacetime or $f\rightarrow \infty$. However, according to \eqref{veflu}, the velocity of the fluid could not be defined for large radius. As an alternative to infinite radius, one could compute the complexity by a finite cut-off surface $r=r_c$ \cite{fcs}. Then, to do the integration in \eqref{co}, one could define a finite cut-off surface. That could be the surface in the fluid with $v=0$, say, near the boundary of the water tank where a fluid is static.}. In addition, Eq. \eqref{co} defines the notion of complexity for fluids analogous to a gravity. In section \ref{realvana}, it will reveal that the extension of some notions in the real black holes to the analogue ones is allowed. Moreover, \eqref{co} is not just a rephrasing of the holographic complexity in terms of the fluid parameters. A relativistic, non-rotational fluid satisfying \eqref{rhoflu}, \eqref{veflu} and \eqref{co} are introducing the fluid system capable of simulating the complexity growth. Put it differently, if one is interested in simulating the interior growth of the black holes\footnote{Complexity studies quantities related to the interior of black holes such as wormholes \cite{comp}} in the fluid and measuring the related quantities, should look at the fluid systems with \eqref{rhoflu}, \eqref{veflu} and \eqref{co} characteristics.

At first glance, it seems that the complexity itself in gravity side is not that useful notion to be measured. However, black holes are holographic duals of strongly-coupled quantum systems such as QGP and some condensed matter systems. In this manner, studying the black holes and measuring complexity could provide some information about the boundary theory and also the original black holes. There are some advantages supporting this idea:
\begin{enumerate}
\item The complexity could be identified as the Lyapunov exponent in cosmology which determines how its dynamics is chaotic \cite{lya, cosmology}. Then by observing the complexity one could measure the chaos of the system. Although the complexity \eqref{co} is defined for a black hole, one can simulate a cosmological background and define the analogous complexity in the model for the observatory purposes.\\
\item The complexity evolves much longer than the system reaches the equilibrium. In other words, when the system macroscopic quantities do not change due to the equilibrium, the complexity does change. This implies that it demonstrates more information than any other quantities in the equilibrium of quantum systems \cite{comp}.\\
\item The complexity could study the geometry of singularity of black holes \cite{prob}. Then measuring the complexity of the analogue black holes could reveal some information about the interior and singularity of the real black holes which are inaccessible to us.
\end{enumerate}

In addition to these advantages, the interpretation of complexity in terms of the analogue gravity brings a considerable benefit to the Lloyd bound. In the following, we show that this interpretation leads to an elucidation of the Lloyd bound in terms of the analogue gravity which relates it to the KSS bound. We have seen that the Lloyd bound in the gravity side constrains the thermodynamic volume in a non-trivial way, then here we expect a non-trivial constraint on fluid quantities. The thermodynamic volume for metric (\ref{ads}) is given by \cite{v}:
\begin{align}
V=\frac{\partial M}{\partial P}=\frac{V_{d-1}R^{d}}{d}\;. \label{tv}
\end{align}
Then complexity growth rate by Eqs. (\ref{pre}), (\ref{l}), \eqref{thermo} and (\ref{tv}) leads to \cite{v}:
\begin{equation}
\dot{C}=\frac{d(d-1)}{16\pi G L^2}.\frac{V_{d-1}R^{d}}{d}= M\Big(1+\frac{k L^2}{R^2}\Big)^{-1}\label{acg}\;,
\end{equation}
and the Lloyd bound implies
\begin{align}
\Big(1+\frac{k L^2}{R^2}\Big)^{-1}&\le 2\;.  \label{comaco} 
\end{align}
For $k=0,1$, the above bound is trivially satisfied. For $k=-1$, we first note that $M\ge 0$ implies $\mu\ge 0$, therefore,
\begin{align}
\mu=R^{d-1}\Big(\frac{R^2}{L^2}+k\Big)\ge 0 \quad\implies\quad R^2\ge L^2\;, \quad \text{for}\quad k=-1\;.
\end{align} 
Then combining with \eqref{comaco}, one finds
\begin{align}
R^2&\ge 2L^2 \;, \quad \text{for}\quad k=-1\;. \label{z0-L}
\end{align}
Eq. \eqref{z0-L} shows that the Lloyd bound is violated in the AdS-Schwarzschild with $k=-1$ for not very large black holes $L^2< R^2< 2L^2$, it means that CV-2.0 proposal is not a proper candidate for the holographic complexity at least in this regime. However, our next results are in the satisfaction domain of $R^2\ge 2L^2$ and are legitimate in this limit.

For experimental convenience, we consider the non-relativistic fluid in the Lloyd bound\footnote{Since the Lloyd bound in CV-2.0 proposal is independent of WDW patch, there would be no confusion about non-relativistic limit.}. By Eqs. (\ref{flux}) and (\ref{z0-L}) it turns into:
\begin{align}
&\Big(\frac{A}{\rho v}\Big)^{\frac{2}{m-1}}\ge -2kL^2\;,\\
& (A)^{\frac{2}{m-1}}\ge (\rho v)^{\frac{2}{m-1}}(-2kL^2)\;.\label{lbn}
\end{align}
We should note that analogue gravity experiments are carried in $3+1$ dimensions so we suppose $m=3$, then the Lloyd bound would be:
\begin{align}
&1\ge \rho v(-2k)(\frac{L^2}{A})\label{lb}\;.
\end{align}
The above inequality that restricts density, velocity, and specific enthalpy of the fluid is analogous to the Lloyd bound or uncertainty principle in the quantum side and could be tested in a condensed matter laboratory as the same as observation of the Hawking radiation in \cite{unr,water}. It can be shown that \eqref{lb} leads to the violation of Kovtun-Son-Starinet (KSS) bound in fluid/gravity duality \cite{son}:
\begin{equation}
\frac{\eta}{s}\ge \frac{1}{4\pi}\label{kss}\;,
\end{equation}
where $\eta$ is the shear viscosity and $s$ is the entropy density of the fluid.  To show that, we write shear viscosity in the fluid as \cite{natsume}:
\begin{equation}
\frac{\eta}{l} \sim \rho v \label{visco}\;,
\end{equation}
where $l$ is the mean free path which by considering a field theory, say $g\phi^4$ at tree level and dimensional analysis, $l$ is proportional to $l\sim 1/(T g^2)$  with $T$ is the temperature and $g$ the coupling constant of the theory \cite{natsume}. By the fact that $s\sim  T^3$ and by the help of \eqref{omeg} in $m=3$, $A\sim T^2$ and $L^2\sim \frac{1}{T^2}$ we find,
\begin{align}
\frac{\rho v}{\frac{A}{L^2}}\sim \frac{\eta/l}{T^4}\sim \frac{\eta Tg^2}{T^4}\sim \frac{\eta g^2}{s}\;,
\end{align}
therefore, we can write \eqref{lb} as follows:
\begin{align}
k=0,1:\quad & \frac{\eta}{s}\ge 0 \label{k0}\;,\\
k=-1:\quad &\frac{\eta}{s}\le \frac{1}{2g^2}\sim\frac{1}{4\pi}\label{k-1}\;.
\end{align}

The inequality \eqref{k0} indeed is trivial and doesn't imply the KSS bound on the viscosity to entropy ratio. In $k=-1$ case, the inequality  \eqref{k-1} strongly violates the KSS bound. These inequalities are the consequence of the uncertainty principle and they strongly violate the KSS bound. There are two assumptions for these inequalities: 1)continuity and Euler equations in fluid dynamics, 2)the Lloyd bound. Put it differently, these inequalities show if the fluid simulates the inner growth of black holes then the KSS bound is violated. However, this is not surprising and is in agreement with some cases for which the KSS bound violation is reported in holography \cite{vio1,vio2,vio3}. Along these lines, it is expected that inequality \eqref{lb} plays a selection rule that verifies the sorts of fluids capable of complexity simulation. In other words, to have a supersonic black hole,  requires $v\ge c$ \cite{a}. Moreover, the inequality \eqref{lb} restricts the growth of the interior of a supersonic black hole or the complexity growth. In the following, we show that the Lloyd bound in the analogue fluid is consistent with the inequalities on the viscosity to entropy ratio when derived from the holography. 

It is worth mentioning that the analogue gravity we have considered yet (the relativistic limit \eqref{am} and the non-relativistic one \eqref{aco}) seems to be a perfect fluid \eqref{energyten}, then as a result this fluid should has no viscosity. However, the viscosity comes into play from the second order perturbation in the energy-momentum tensor and the second order does not affect the analogue metrics; in other words, the analogue metrics are shared between inviscid and viscose fluids and the higher order perturbations in the energy-momentum of the fluid are accommodated to the field theories, say $\phi$ in \eqref{lagrang}, on the curved spacetime \cite{viscos}. In a layman's term, the perfect fluid plays the role of the background and the perturbation around the perfect fluid, the role of the field theory fluctuation.

\subsection{AdS Gauss-Bonnet}
It is known that the Gauss-Bonnet (GB) gravity is where the KSS bound is violated \cite{gbv,vio1,gbv2,gbv3,gbv4}. However, we should note that in the next section we demarcate two fluids corresponded to a gravity theory, a boundary fluid which is from fluid-gravity duality and the other from the analogue gravity. The KSS bound is violated in the boundary fluid dual to the GB gravity and we do not suppose that it violates in the analogue gravity. Regarding the relation between the KSS bound and the Lloyd bound, it would be interesting to verify independently, if the Lloyd bound is violated in the analogue fluid for the GB gravity. However, the first step would be looking for an analogue for the GB gravities, if any, and investigating whether it fulfills the continuity and Lagrange equations. A neutral AdS-Gauss-Bonnet (AGB) is given by:
\begin{align}
&S=\frac{1}{16\pi G} \int d^{d+1}x \sqrt{-g} \Big(R-2\Lambda+\alpha (R^2-4R_{\mu\nu}R^{\mu\nu}+R_{\mu\nu\rho\sigma}R^{\mu\nu\rho\sigma})\Big) \nonumber \;,\\
&ds^2=-f(r)dt^2+\frac{dr^2}{f(r)}+r^2 d\sigma_{d-1}^2 \label{gb}\;,\\
&f(r)=k+\frac{r^2}{2\lambda L^2}\Big(1\pm \sqrt{1+\frac{64\pi \lambda L^2}{d-1} \big(\frac{M}{r^d\sigma_{d-1}}-\frac{P}{d}\big)}\Big)\label{bla}\;,
\end{align}
and for $``-"$ sign, the theory continuously goes to Einstein-AdS solution in limit $\alpha \rightarrow 0$. $L$ is the AdS radius, $\lambda=\alpha (d-2)(d-3)/L^2$, and curvature parameter $k$ could be $1, 0$ and $-1$ in spherical, planar and hyperbolic horizon topology, respectively. The thermodynamic quantities of the black hole is read as:
\begin{align}
&M=\frac{(d-1)\sigma_{d-1}R^{d-2}}{16\pi G}\Big(k+\frac{k^2\lambda L^2}{R^2}+\frac{16\pi P R^2}{d(d-1)}\Big)\;,\\
&V=\frac{\sigma_{d-1}R^{d}}{d}\;,\\
&P=-\frac{\Lambda}{8\pi G}=\frac{d(d-1)}{16\pi G L^2}\;.
\end{align}
The complexity growth rate would be:
\begin{align}
&\dot{C}=PV\le2M\;,\\
&\frac{d(d-1)}{16\pi G L^2}.\frac{\sigma_{d-1}R^{d}}{d}\le 2\;.\frac{(d-1)\sigma_{d-1}R^{d-2}}{16\pi G}\Big(k+\frac{k^2\lambda L^2}{R^2}+\frac{16\pi P R^2}{d(d-1)}\Big)\;,\\
&\frac{1}{L^2}\le \frac{2}{R^2}\Big(k+\frac{k^2\lambda L^2}{R^2}+\frac{R^2}{L^2}\Big)\;,\\
&0\le \frac{2}{R^2}\Big(k+\frac{k^2\lambda L^2}{R^2}+\frac{R^2}{2L^2}\Big)\;,\\
&0\le k+\frac{k^2\lambda L^2}{R^2}+\frac{R^2}{2L^2}\label{gbie}\;.
\end{align}

To make sure \eqref{gb} has an analogue, we add a conformal factor as the same as what we did in \eqref{conformal}:
\begin{align}\label{egbml}
ds^2=\Omega^2\Big(-f(r)dt^2+\frac{dr^2}{f(r)}+r^2 d\sigma_{d-1}^2\Big)\;.
\end{align}
By change of variable $t= \tau+\int \frac{\sqrt{1-f(r)}}{f(r)}dr$, the line element \eqref{egbml} in Painleve-Gulltstrand coordinates could be rewritten as:
\begin{equation}
ds^2=\Omega^2\Big(-f(r)  d\tau^2+dr^2-2\sqrt{1-f(r)} d\tau dr+r^2 d\sigma_{d-1}^2\Big) \label{egbmp}\;.
\end{equation}
Comparing \eqref{egbmp} with \eqref{aco} we arrive at \eqref{ro} and \eqref{ve}:
\begin{align}
&\rho=\mathcal{K}c \Omega^{m-1}\label{ro}\;,\\
&v=\sqrt{1-f}\label{ve}\;,\\
&\rho v=\frac{L^2}{r^{m-1}}\;,
\end{align}
the flux at the horizon in $m=3$ is given by:
\begin{align}
\frac{R^{2}}{L^2}=\frac{1}{\rho v}\sim \frac{s}{\eta g}\;,
\end{align}
then \eqref{gbie} by the flux at the horizon is read:
\begin{align}
&0\le k+\frac{k^2\lambda}{\frac{s}{\eta g}}+\frac{s}{2\eta g}\;,\\
&0\le 2 gk\frac{\eta}{s}+2 g^2 k^2\lambda\frac{\eta^2}{s^2}+1\;.
\end{align}
For $k=0,1$, the above inequality is satisfied and leads to $\frac{\eta}{s}\ge 0$. For $k=-1$, it is satisfied either with $2\lambda>1$ or we have,
\begin{align}
&\frac{\eta}{s}\ge \frac{1+\sqrt{1-2\lambda}}{2g\lambda}\;, \quad \text{or}\quad \frac{\eta}{s}\le \frac{1-\sqrt{1-2\lambda}}{2g\lambda} \label{gbkss}\;.
\end{align}
The left inequality in \eqref{gbkss} respects the KSS bound. Due to the fact that the Lloyd bound in some cases respects the KSS bound and in some cases violates it, and the fact that there are some evidences in holography that confirm the violation of the KSS bound.

\section{Acoustic Black Hole Vs. Real Black Hole} \label{realvana}
In previous sections, by transformations (\ref{roo}) and (\ref{vee}), we made an acoustic black brane and computed complexity for an acoustic black hole. However, there is a difference between a real black brane and its acoustic counterpart \cite{ra}.
Besides that analogue gravity provides an analogy to correspond a gravity theory to a fluid which we call it {\it analogue fluid}, AdS/CFT makes a one-to-one correspondence between gravity and fluid which we call it {\it boundary fluid}. It means that gravity in the $D+1$ dimensional bulk corresponds to a $D$ dimensional fluid on the boundary of the bulk \cite{fg}. In this manner, we call a black hole in the bulk a real black hole. One can ask how the analogue fluid and the boundary fluid are related. The answer is that if we have fluid on a time-like cut-off $r=r_c$, when the cut-off goes to the boundary of the bulk, the analogue fluid and the boundary fluid coincide \cite{ra}. Analogue gravity on a cut-off surface is one of the representation of the analogue black hole in terms of original black holes. However, this is a different interpretation from the one in section \ref{review} where it has been shown that the realization of the analogue black holes could be in general spacetime backgrounds. It means that in laboratories with flat spacetime, one could simulate all sort of original black holes, an arbitrary background. In contrast, this new interpretation is desirable when it comes to the comparison between the holographic result and the analogue gravity one. In the following, we consider SVER in either holographic and analogue gravity manners and see how the results are replicated.

We want to show how the complexity growth rate of a real black hole and an acoustic one are related. Suppose that we have a black brane in a $D+1$ dimensional bulk:
\begin{equation}
ds^2=\frac{L^2}{r^2f}dr^2+\frac{r^2}{L^2} (-f dt^2+d\vec{x}^2)\;,
\end{equation}
where
\begin{align*}
f=1-\Big(\frac{r_h}{r}\Big)^D \;.
\end{align*}
and $r_h$ is the horizon radius. The temperature of the black brane and the fluid on the boundary read as:
\begin{equation} 
T_B=\frac{D~r_h}{4\pi L^2}\,.
\end{equation}
The complexity growth rate for this black brane would be:
\begin{align}
&\dot{C}=PV=\frac{-\Lambda}{8\pi G_{D+1}}\frac{l^{D-1} r_h^D}{D }=\frac{-\Lambda}{8\pi G_{D+1}}(4\pi)^D ~l^{D-1}L^{2D}\Big(\frac{1}{D}\Big)^{D+1}~T_B^D \label{rco}\;,\\
&\Lambda=-\frac{D(D-1)}{2 L^2}\;,\\
&l^{D-1}=V_{D-1}=\int d^{D-1}x\;.
\end{align}

By finding entropy of the black brane and using the thermodynamics of black holes, energy density and the pressure of the fluid on the boundary (fluid boundary) are in hand. Then using renormalized holographic energy tensor at cut-off surface $r=r_c$, the pressure and the energy density of the fluid at the cut-off surface could be computed. Put it another way, the analogue gravity (analogue fluid) at the cut-off surface is obtained \cite{ra}:
\begin{align}
&ds^2=\Big(\frac{n_c^2}{c_s (\rho_c+p_c)}\Big)^{\frac{2}{D-2}}\Big(\eta_{\mu\nu}+(1-c_s^2)\tilde{u}_{\mu}\tilde{u}_{\nu}\Big)dx^{\mu}dx^{\nu} \label{afg}\;,\\
&c_s^2=\frac{1}{D-1}+\frac{D r_h^D}{2(D-1)r_c^Df(r_c)} \label{ss}\;,\\
&\tilde{u}_\mu=\sqrt{f(r_c)}u_\mu \;,
\end{align}
where subscript $c$ denotes quantities at the cut-off surface. If one compares Eq. \eqref{afg} with Eq. \eqref{am}, realizes that $D=m+1$. Then, computations in section \ref{spr} are repeated here. To be more concrete, it should be noted that Eqs. (\ref{roo}) and (\ref{vee}) mean that the analogue fluid behaves as a black brane in AdS space (\ref{ads}). As mentioned in the preceding section, there is no guarantee that whether a gravitational theory has an analogue model, then here we suppose that for a $d+1$ gravitational theory there is a $D$ dimensional fluid. Hence, as in the previous section, we compute complexity of a $d+1$ gravitational theory and rewrite it in terms of quantities in a $D$ dimensional fluid.
Then Eqs. (\ref{roo}) and (\ref{vee}) turn into:
\begin{align}
&\rho=\frac{L^{D-2}\mathcal{K}c_s}{z^{D-2}} \label{aro}\;,\\
&\tilde{v}^2=f(r_c) v^2=\Big(\frac{z}{z_0}\Big)^d \label{ave}\;.
\end{align}
Eq. (\ref{ave}) by use of (\ref{aro}) can be written:
\begin{equation}
z_0=\frac{L \Big(\frac{\mathcal{K}c_s}{\rho}\Big)^{\frac{1}{D-2}}}{\Big(\frac{f(r_c)v^2}{c_s^2}\Big)^{\frac{1}{d}}}=\frac{L \Big(\frac{\mathcal{K}}{\rho}\Big)^{\frac{1}{D-2}}}{\Big(f(r_c)v^2\Big)^{\frac{1}{d}}}~c_s^{\frac{1}{D-2}} \label{z0}\;.
\end{equation}
Just similar to section \ref{spr}, the complexity growth rate for the analogue fluid (\ref{afg}) is given by (\ref{acg}):
\begin{align*}
\dot{C}=\frac{d(d-1)}{16\pi G_{d+1} L^2}.\frac{V_{d-1}R^{d}}{d}=\frac{V_{d-1}(d-1) L^{2d-2}}{16\pi G_{d+1} z_0^{d}}:=\dot{C}_c\;,
\end{align*}
that could be rewritten by (\ref{z0}) as follows:
\begin{equation}
\dot{C}_c=\frac{V_{d-1}(d-1) L^{2d-2}}{16\pi G_{d+1} \bigg(\frac{L \Big(\frac{\mathcal{K}}{\rho}\Big)^{\frac{1}{D-2}}}{\Big(f(r_c)v^2\Big)^{\frac{1}{d}}}~c_s^{\frac{1}{D-2}}\bigg)^{d}}=\frac{l^{d-1}(d-1) L^{2d-2}}{16\pi G_{d+1} \bigg(\frac{L \Big(\frac{\mathcal{K}}{\rho}\Big)^{\frac{1}{D-2}}}{\Big(f(r_c)v^2\Big)^{\frac{1}{d}}}~c_s^{\frac{1}{D-2}}\bigg)^{d}}\label{aic}\;.
\end{equation}
The horizon radius in terms of complexity of the real black hole (\ref{rco}) is given by:
\begin{equation}
r_h^D=\frac{16\pi G_{D+1} L^2~\dot{C}}{l^{D-1}(D-1)}\label{hr}\;,
\end{equation}
by (\ref{ss}) and (\ref{hr} ) in hand, complexity of analogue gravity (\ref{aic}) reads as:
\begin{align}
&\dot{C}_c=\frac{l^{d-1}(d-1) L^{2d-2}}{16\pi G_{d+1} \bigg(\frac{L \Big(\frac{\mathcal{K}}{\rho}\Big)^{\frac{1}{D-2}}}{\Big(f(r_c)v^2\Big)^{\frac{1}{d}}}~\Bigg[\frac{1}{D-1}+\frac{D \Big(\frac{16\pi G_{D+1} L^2~\dot{C}}{l^{D-1}(D-1)}\Big)}{2(D-1)r_c^Df(r_c)}\Bigg]^{\frac{1}{2(D-2)}}\bigg)^{d}}\;,\\
&\dot{C}_c=\frac{l^{d-1}(d-1) (D-1)^{\frac{d}{2(D-2)}}L^{d-2}  v^{2}}{16\pi G_{d+1} \Big(\frac{\mathcal{K}}{\rho}\Big)^{\frac{d}{D-2}} \Bigg[f(r_c)+\frac{ 8\pi D G_{D+1}L^2~\dot{C}}{l^{D-1}(D-1)r_c^D}\Bigg]^{\frac{d}{2(D-2)}}} f(r_c)^{1+\frac{d}{2(D-2)}}\label{realandaco}\;,
\end{align}

we call it {\it complexity growth rate of analogue fluid} at the cut-off surface. Eq. \eqref{realandaco} defines the complexity growth rate of an acoustic black hole $\dot{C}_c$ in terms of the complexity growth rate of a real black hole $\dot{C}$. If we suppose that the fluid which we conduct an experiment on it, is a $3+1$ dimensional fluid ($D=3+1$) and the gravitational theory that is simulated partly is a $4+1$ dimensional ($d=4$ as we did in the preceding section), then Eq. \eqref{realandaco} is written as:
\begin{align}
\dot{C}_c=\frac{9 l^{3}L^6  v^2}{16\pi G_{5} \Big(\frac{\mathcal{K}}{\rho}\Big)^2 \Bigg[f(r_c)+\frac{ 32\pi G_{5}L^2~\dot{C}}{3 l^{3}r_c^4}\Bigg]} f^2(r_c) \label{2com}\;.
\end{align}
This relation shows that one can measure the complexity growth rate of the analogue gravity (analogue fluid) $\dot{C}_c$ in a laboratory and find out the complexity growth rate of a real black hole. The idea provides a simulation procedure because for finding the complexity of a quantum circuit, one can equivalently measure a quantity (\textit{complexity of the fluid}) in a condensed matter laboratory such as a water tank \cite{water} and figure out what the complexity of the quantum circuit is. The same procedure could be conducted for the entanglement entropy.

Eq. \eqref{2com} provides a relation between the mass of the real black hole and that of the acoustic black hole. By the fact that in real black hole $\dot{C}\le 2M$ and ,by definition, in acoustic black hole $\dot{C}_c\le 2M_c$, eq.\eqref{comaco}, we can use \eqref{2com} as follows \footnote{We set $L=G=1$}:

\begin{align}
&\dot{C}_c=\frac{9 l^{3}  v^2}{16\pi  \Big(\frac{\mathcal{K}}{\rho}\Big)^2 \Bigg[f(r_c)+\frac{ 32\pi ~\dot{C}}{3 l^{3}r_c^4}\Bigg]} f^2(r_c)\le 2M_c\;,\\
&\frac{9l^3v^2 f^2(r_c)}{2M_c 16\pi \Big(\frac{\mathcal{K}}{\rho}\Big)^2 }  \le \Bigg[f(r_c)+\frac{ 32\pi ~\dot{C}}{3l^3 r_c^4}\Bigg]\;,\\
&\frac{9l^3v^2 f^2(r_c)}{2M_c 16\pi \Big(\frac{\mathcal{K}}{\rho}\Big)^2 } -f(r_c) \le \frac{ 32\pi ~\dot{C}}{3l^3 r_c^4}\;,\\
&\frac{3l^3 r^4_c}{32\pi}\Bigg(\frac{9l^3v^2 f^2(r_c)}{2M_c 16\pi \Big(\frac{\mathcal{K}}{\rho}\Big)^2 } -f(r_c)\Bigg) \le \dot{C}\le2M\;,\\
&\frac{3l^3 r^4_c}{64\pi}\Bigg(\frac{9l^3v^2 f^2(r_c)}{32\pi M_c \Big(\frac{\mathcal{K}}{\rho}\Big)^2 } -f(r_c)\Bigg)\le M \label{mass}\;.
\end{align}
Eq. \eqref{mass} which is based on the renormalization holography and the definition \eqref{acg}, restricts the mass of the real black hole and that of acoustic.

Besides the relation between the complexity of the two fluids, it could be shown that there is a relation between the quantities $\eta/s$ in the two fluids. For a general solution of the following form:
\begin{align}
ds^2=2dtdr -\bar{F}(r)dt^2+\bar{G}(r)dx^2\label{gensol}\;,
\end{align}
by rescaling the coordinates $t'=\sqrt{\bar{F}(r_c)}t$, $x'=\sqrt{\bar{G}(r_c)}x$ and boosting the metric, one finds \cite{general}:
\begin{align}
ds^2=-\frac{2u_idx'^idr}{\beta}+\Big(G(r)-F(r)\Big)(u_idx'^i)^2+G(r)dx'_idx'^i\;,
\end{align}
where $F(r):=\frac{\bar{F}(r)}{\bar{F}(r_c)}$, $G(r):=\frac{\bar{G}(r)}{\bar{G}(r_c)}$, $\beta:=\sqrt{\bar{F}(r_c)}$, $u^i=\gamma (1,v^i)$ and $\gamma=\frac{1}{\sqrt{1-v^2}}$. To find shear viscosity one perturbs the metric and uses the following decomposition:
\begin{align}
g^{(n)}_{\mu\nu}=\phi^{(n)}u_{\mu}u_{\nu}+2u_{(\mu}\psi^{(n)}_{\nu)}+H^{(n)}_{\mu\nu}+\frac{1}{d}H^{(n)}P_{\mu\nu}\label{decom}\;,
\end{align}
where the projection $P_{\mu\nu}=\eta_{\mu\nu}+u_{\mu}u_{\nu}$. Then for the analogue fluid (on the cut-off surface) we get into the result \cite{general}:
\begin{align}
\frac{\eta(r_c)}{s(r_c)}\sigma_{\mu\nu}=\frac{\frac{1}{2}\beta P_{\mu\lambda}P_{\nu\gamma}H'^{(1)\lambda \gamma} (r_c)+\sigma_{\mu\nu}}{4\pi G(R)^{\frac{d}{2}}}:=\frac{\mathcal{H}_{\mu\nu}(r_c)}{4\pi}\;,
\end{align}
where $\sigma_{\mu\nu}=P^i_{(\mu}P^j_{\nu)}\partial_iu_j-\frac{1}{d}P_{\mu\nu}\partial_iu^i$. By the fact that $\frac{\eta}{s}\Big|_{boundary}=\lim_{r_c\to \infty}\frac{\eta(r_c)}{s(r_c)}$, for any non-vanishing component of $\sigma_{\mu\nu}$:
\begin{align}
\frac{\eta(r_c)}{s(r_c)}=B(r_c)\frac{\eta}{s}\Big|_{boundary} \qquad\text{with}\qquad B(r_c):=\Big(\frac{\mathcal{H}_{\mu\nu}(r_c)}{\mathcal{H}_{\mu\nu}(\infty)}\Big)\label{etas}\;,
\end{align}
this relation shows that $\eta/s$ of the two fluids are proportional. In appendix \ref{app1}, we derive $B(r_c)$ numerically and show that it is always less than one,  $B(r_c)<1$.

Now we can get back to the inequalities in the previous section. \eqref{k-1} which is the Lloyd bound of the Schwarzschild analogue for $k=-1$ leads to the violation of the KSS bound, in other words \eqref{k-1} is the quantity for the analogue fluid:
\begin{align}\label{lloyd-k-1}
\frac{\eta(r_c)}{s(r_c)}\le \frac{1}{2g}\;,
\end{align}
where $g$ the coupling constant rests on the location of the cut-off surface, then $g=g(r_c)$. By using \eqref{etas}:
\begin{align}
\frac{\eta}{s}\Big|_{boundary}\le \frac{1}{2g}\frac{1}{B(r_c)}\label{scbo}\;.
\end{align}
This is the Lloyd bound imposed on the boundary fluid. 

On the other hand, $\eta/s$ for the boundary fluid can be calculated directly from holography and is found to be \cite{kss-1}:
\begin{align}
\frac{\eta}{s}\Big|_{boundary}= \frac{1}{4\pi}\Big(1-\frac{1}{2r_+^2}\Big)^2\le \frac{1}{4\pi}\;.
\end{align}
This shows that the holographic calculation gives a result which is consistent with the Lloyd bound in \eqref{lloyd-k-1} and \eqref{scbo}\footnote{Note that as approaching the boundary, the bound $1/(2g)$ can be replaced with $1/(4\pi)$ and recall that we have $B(r_c)<1$ as shown in appendix \ref{app1}.}.  

The same result appears in GB gravity. The second inequality in \eqref{gbkss} with $2\lambda\ll 1$ leads to:
\begin{align}
\frac{\eta(r_c)}{s(r_c)}\ge \frac{2-\lambda}{2g\lambda}\approx\frac{1}{g\lambda}\quad\text{or}\quad\frac{\eta(r_c)}{s(r_c)}\le \frac{1}{2g}\;.
\end{align}
Then for the boundary fluid, using \eqref{etas}, we find:
\begin{align}\label{GBloyd-boundary}
\frac{\eta}{s}\Big|_{boundary}\ge \frac{1}{2g\lambda}\frac{1}{B(r_c)} \quad\text{or}\quad \frac{\eta}{s}\Big|_{boundary}\le \frac{1}{2g}\frac{1}{B(r_c)}\;.
\end{align}
These are necessary conditions derived by imposing the Lloyd bound on the analogue fluid with small $\lambda$.  

However, the holography reveals that the quantity on the boundary is given by \cite{gbboundary}
\begin{align}
\frac{\eta}{s}\Big|_{boundary}= \frac{1}{4\pi}\Big(1-4\lambda\Big)\le \frac{1}{4\pi}\label{gbboundary}\;.
\end{align}
It seems that if we take the holographic calculations as a presumption, say, \eqref{gbboundary}, then the second inequality in \eqref{GBloyd-boundary} is satisfied and the Lloyd bound follows. But the converse does not hold necessarily. It means that from \eqref{GBloyd-boundary} we can not necessarily infer \eqref{gbboundary}. 
Put it differently, given the holographic result for the boundary SVER, and according to \eqref{etas}, if the boundary SVER violates the KSS bound, then the bulk SVER ($\eta(r_c)/s(r_c)$) violates too. However, the inverse is not true, i.e., if the boundary SVER satisfies the KSS bound, then the bulk SVER could either satisfy the KSS bound or not. In other words, the satisfaction of the KSS bound on the boundary is a necessary condition for the satisfaction of its bulk counterpart.

\section{Conclusion}
Recently the notion of computational complexity has been introduced for black holes through AdS/CFT correspondence. The computational complexity provides information about the interior of black holes. However, black holes and some other gravitational phenomena are not available yet to conduct experiments on them. Although, there are some opportunities in PBH to detect Hawking radiation, these possibilities do not come up with the accessibility to the interior of black holes. Nonetheless, analogue gravity provides some techniques to simulate gravitational theories with condensed matter systems, and vice versa, simulating condensed matter systems with gravitational phenomena. These analogue systems have the advantage of observing some gravitational phenomena that are not directly available in the original gravitational experiments such as evaporation of black holes, can be simulated in condensed matter laboratories such as acoustic black holes. 

There are some suggested proposals for holographic computational complexity. These proposals are performed by numerical computations but among them, the CV-2.0 proposal can be proceeded by more analytical computations rather than numerical, especially for the complexity growth rate. In this work, we have studied computational complexity by CV-2.0 proposal which relates spacetime volume restricted to WDW patch to complexity. By using acoustic black holes formed in an irrotational-relativistic fluid, we found an interpretation of complexity in condensed matter systems which is related to the volume integration of the flux density of the fluid Eq. (\ref{co}) where our results are a simulation of a $d+1$ dimensional gravitational theory with a $m+1$ dimensional fluid. 

The analogue interpretation of the complexity and measuring it in laboratories gains the upper hand in some cases such as:  
\begin{itemize}
\item Measuring the chaos in cosmology through the complexity (which is equal to the Lyapunov exponent).\\
\item Considering the condensed matter systems long after their equilibrium. Measuring the complexity provides information about the microstates of the system.\\
\item Study the geometry of black holes singularities which are directly inaccessible to us forever.
\end{itemize}

We also studied the Lloyd bound which is originated from the uncertainty principle and constraints the complexity growth rate. We found that there would be a non-trivial constraint on the fluid flux $\rho v$ as given in Eq. (\ref{lb}). This constraint in the fluid is rooted in the uncertainty principle. To do that, we restricted ourself to the non-relativistic limit for the matter of experimental convenience. To this extent there is no confusion on the non-relativistic limit of the Lloyd bound and the WDW patch in definition of the complexity \eqref{co}, because the Lloyd bound in CV-2.0 is proportional to the thermodynamic volume of the black hole instead of the volume of the spacetime in WDW patch. We showed that the Lloyd bound is reduced to the shear viscosity to entropy ratio (SVER). As a consequence, the Lloyd bound or SVER could satisfy the KSS bound or violate it. AdS-Schwarzschild black hole for any kinds of horizon geometries violates the KSS bound and AdS-Guass-Bonnet black hole violates the KSS bound in $k=0,1$ and in $k=-1$ there would be two inequality, one of them respects the KSS bound and the other violates it, but to be consistent with the holographic result we should take the violating one. However, the Lloyd bound could be violated only in AdS-Schwarzschild $k=-1$ for black holes horizon smaller than AdS radius \cite{v}.

Moreover, we showed that although it is not trivial to develop the notion of computational complexity to a fluid, the substitution of fluid quantities in the Lloyd bound leads to the SVER which is consistent with the holographic computation. For this purpose, we used another interpretation of the analogue gravity which is located on a time-like cut-off surface of an asymptotic AdS bulk where on the boundary the fluid in fluid-gravity duality lives. This new interpretation allows comparison between the analogue gravity results and the holographic ones and \eqref{etas} shows that they are related by a $B$ function which is less than one, then the SVER in the analogue gravity is greater or equal to the SVER in the boundary fluid.
We compared SVER of the analogue fluid and the boundary fluid in AdS-Schwarzschild and AdS-Gauss-Bonnet gravity for the hyperbolic black holes ($k=-1$) as shown in the following table.

\renewcommand{\arraystretch}{2}
\begin{center}
\begin{tabular}{ |c|c|c| } 
 \hline
\textbf{} & \textbf{AdS-Schwarzschild}&\textbf{AdS-GB} \\
\hline  
 \textbf{Boundary fluid} & $\frac{\eta}{s}\Big|_{boundary}= \frac{1}{4\pi}\Big(1-\frac{1}{2r_+^2}\Big)^2\le \frac{1}{4\pi}$ & $\frac{\eta}{s}\Big|_{boundary}= \frac{1}{4\pi}\Big(1-4\lambda\Big)\le  \frac{1}{4\pi} $ \\
\textbf{Analogue fluid} & $\frac{\eta(r_c)}{s(r_c)}\le \frac{1}{2g}$ & $\frac{\eta(r_c)}{s(r_c)}\ge \frac{1}{g\lambda}~~\text{or}~~\frac{\eta(r_c)}{s(r_c)}\le \frac{1}{2g}$ \\
 \hline
\end{tabular}
\end{center}
\renewcommand{\arraystretch}{1}

In a nutshell, it has been revealed that the satisfaction of the KSS bound in the boundary fluid is a necessary condition for the satisfaction of the KSS bound in the bulk fluid (the analogue fluid). In other words, whenever the KSS bound is violated in the boundary fluid, the KSS bound is violated in the analogue fluid. In addition, when the KSS bound is respected in the boundary fluid, the KSS bound could be either satisfied or violated in the analogue fluid.

Finally, the interpretation of the Lloyd bound brings the information theory (the Lloyd bound) into the analogue gravity. As a consequence, one could investigate the famous information paradox in evaporation of black holes in the analogue gravity parlance. In \cite{infoana}, we suggested that the information loss is interpreted as a momentum loss over the horizon of the analogue black hole and the resolution of the information paradox as a maintenance of the momentum to satisfy the Newton's second law. This analogue gravity interpretation provides a non-ad hoc resolution to the information paradox instead of an ad hoc resolution such as the island prescription. All in all, it seems that the analogue gravity could yield to a better understanding of the physical phenomena or at least shed new light on the physics problems.

\section*{Acknowledgment}
We would like to appreciate Ahmad Moradpoori and Mehdi Sadeghi for their useful comments and discussions.

\appendix
\section{Viscosity to entropy ratio in Gauss-Bonnet Gravity}\label{app1}
Here we try to derive the viscosity to entropy ratio in the analogue fluid from that of the boundary fluid in the case of GB gravity with hyperbolic black hole, $k=-1$. Specifically, we are going to derive the coefficient $B(r_c)$ in \eqref{etas}. We need to solve the Einstein equation for the decomposition \eqref{decom} and find $H_{\mu\nu}^{(n)}$. For general solution \eqref{gensol} containing $H_{\mu\nu}^{(n)}$ Einstein eq. leads to \cite{general}\footnote{Without loss of generality and for the mater of convenience, we do the calculations in $d+1=4$. In $d+1=4$ the energy tensor is reduced to the AdS background \cite{etgb}}:
\begin{align}\label{h-equation}
-\frac{\Gamma^2}{2}F\partial_r^2 H_{ab}^{(n)}-\frac{\Gamma^2}{2}\Big(\frac{d-5}{2}\frac{FG'}{G}+F'\Big)\partial_r H_{\mu\nu}^{(n)}-\frac{\Gamma^2}{2}\frac{FG'^2}{G^2}H_{\mu\nu}^{(n)}\nonumber\\
+\Big(R_{cd}^{(n)}-T_{cd}^{(n)}\Big)\Big(P_a^cP_b^d-\frac{1}{d-1}P^{cd}P_{ab}\Big)=0\;,
\end{align}
in which
\begin{align}
& T_{cd}^{(1)}\Big(P_a^cP_b^d-\frac{1}{d-1}P^{cd}P_{ab}\Big)=\frac{2}{d-1}\Lambda H_{cd}^{(1)}\;,\\
&R_{cd}^{(1)}\Big(P_a^cP_b^d-\frac{1}{d-1}P^{cd}P_{ab}\Big)=\frac{d-1}{2}G'\sigma_{ab}\;,
\end{align}
and where $F$ and $G$ are metric components in Eddington-Finkelstein coordinates, so in GB gravity compared with \eqref{gensol}, they are given as,
\begin{align}
&G=\frac{\bar{G}}{\bar{G}(r_c)}=\frac{r^2}{r_c^2}\;,\\
&F=\frac{r^2}{r_c^2}\frac{f(r)}{f(r_c)}\;,\\
&f(r)=k+\frac{r^2}{2\lambda L^2}\Big(1- \sqrt{1+\frac{64\pi \lambda L^2}{d-1} \frac{M}{r^d\sigma_{d-1}}}\Big)\;,\\
&\Gamma=\frac{r_c}{L}\sqrt{f(r_c)}\;.
\end{align}

We can change variable to $z=r_c/r$ with $0\leq z\leq 1$ and try to solve for $h(z)$ where $H_{ab}=h(z)\sigma_{ab}$. The near boundary solution ($z\to 0$) is found to be,
\begin{align}\label{asymp}
h(z)\approx \frac{4 \lambda  \sqrt{3-320 \pi  \lambda }}{\left(320 \sqrt{3} \pi  \lambda +3 \sqrt{3-320 \pi  \lambda }-3 \sqrt{3}\right)r_c^3} z +\mathcal{O}(z^3)\;.
\end{align}
where we take $\lambda\leq 3/(320\pi)\approx 0.00298$. 

\begin{figure}
	\captionsetup{width=0.8\textwidth}
	\begin{center}
		\includegraphics[height=55mm]{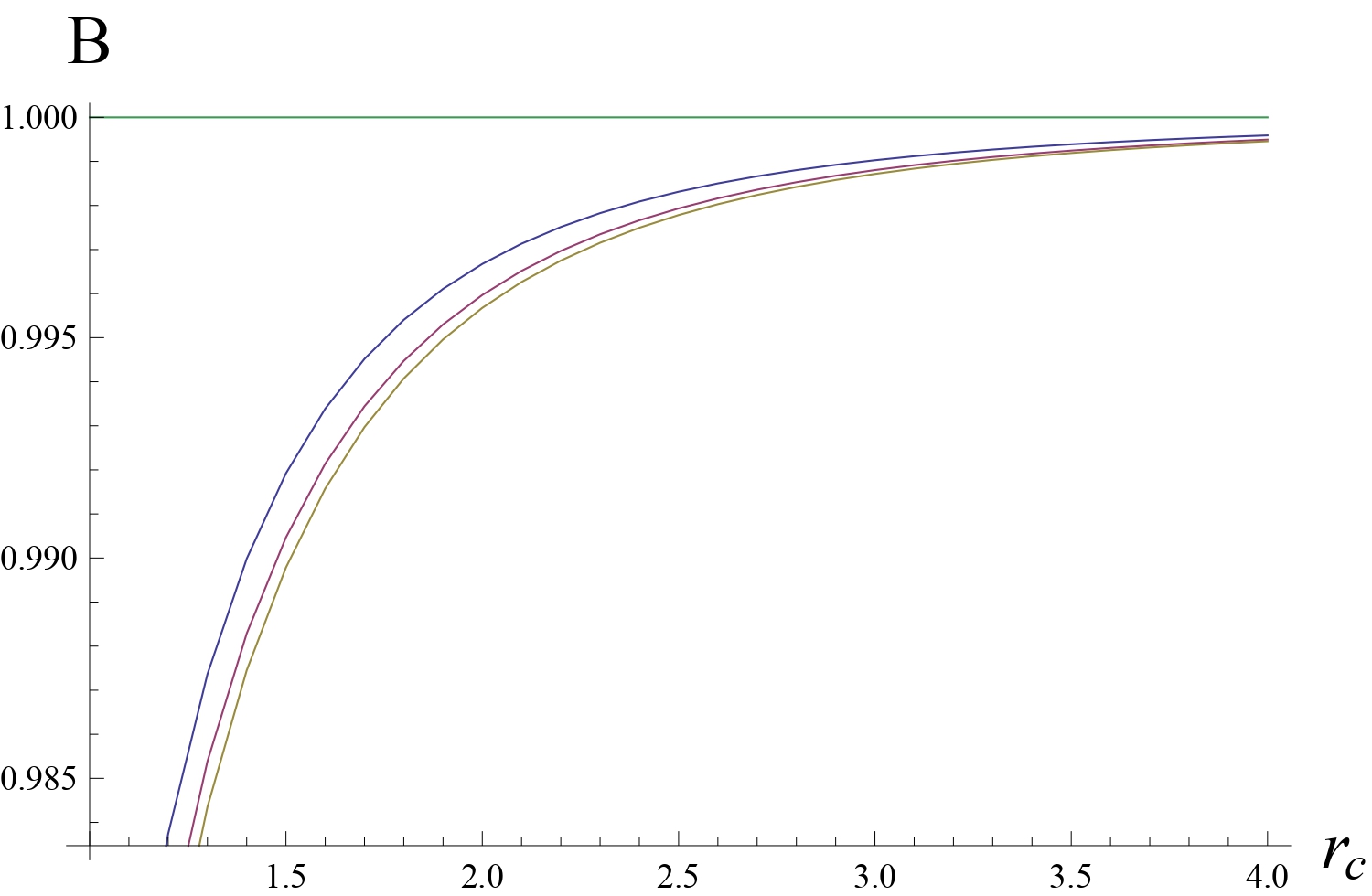} \\
		\caption{$B(r_c)$ as a function of $r_c$ for $d=4$. Curves from bottom to top correspond to $\lambda=.001, .002, .00298$. All of them are always below constant line 1.  }\label{fig:B}	
	\end{center}
\end{figure}

Now from the asymptotic solution \eqref{asymp}, we can read the initial condition for $h(0)$ and $h'(0)$ and solve equation \eqref{h-equation} numerically. We take $\lambda=0.002$ and find the factor $B(r_c)$ in \eqref{etas} for different values of $r_c$ as depicted in Fig. \ref{fig:B}. It shows that $B(r_c)<1$ for all $r_c$.   
\\


\begin{thebibliography}{99}
\bibitem{mal} J. Maldacena, ''The Large N Limit of Superconformal Field Theories and Supergravity," Adv.Theor.Math.Phys.2:231-252,1998, arXiv:9711200 [hep-th].
\bibitem{comp} L. Susskind, "Three Lectures on Complexity and Black Holes,"	arXiv:1810.11563 [hep-th].
\bibitem{ca}A. R. Brown, D. A. Roberts, L. Susskind, B. Swingle and Y. Zhao, "Complexity Equals Action," Phys. Rev. Lett. 116, 191301 (2016), 	arXiv:1509.07876[hep-th].
\bibitem{shock} L. Susskind and D. Stanford, "Complexity and shock wave geometry," Phys. Rev. D 90, 126007 (2014), arXiv:1406.2678 [hep-th].
\bibitem{couch}J. Couch, W. Fischler, P. H. Nguyen, ''Noether charge, black hole volume, and complexity," JHEP 1703 (2017) 119, arXiv:1610.02038 [hep-th].
\bibitem{lloyd} S. Lloyd, "Ultimate physical limits to computation," Nature 406 (Aug., 2000), arXiv:9908043[quant-ph].
\bibitem{myers}D. Carmi, S. Chapman, H. Marrochio, R. C. Myers and S. Sugishita, "On the Time Dependence of Holographic Complexity," JHEP 1711 (2017) 188, arXiv:1709.10184 [hep-th].
\bibitem{Moosa} M. Moosa, "Divergences in the rate of complexitication," arXiv:1712.07137 [hep-th].
\bibitem{Ghodrati} M. Ghodrati, "Complexity growth in massive gravity theories, the effects of chirality, and more," Phys. Rev. D 96, no. 10, 106020 (2017).
\bibitem{Yang} R. Q. Yang, C. Niu, C. Y. Zhang and K. Y. Kim, "Comparison of holographic and field theoretic complexities for time-dependent thermofield double states," JHEP 1802, 082 (2018).
\bibitem{alish}M. Alishahiha, A. Faraji Astaneh, M. R. Mohammadi Mozaffar and A. Mollabashi, "Complexity Growth with Lifshitz Scaling and Hyperscaling Violation," JHEP 1807 (2018).
\bibitem{lc} Z.-Y. Fan and M. Guo,"On the Noether charge and the gravity duals of quantum complexity," JHEP08(2018)031, arXiv:1805.03796v2 [hep-th].


\bibitem{gravitywave}R. Dong, W. H. Kinney and D. Stojkovic, "Gravitational wave production by Hawking radiation from rotating primordial black holes,"  JCAP10(2016)034,  arXiv:1511.05642v3 [astro-ph.CO].
\bibitem{xray}G. Ballesteros, J. Coronado-Blázquez and D. Gaggero
, "X-ray and gamma-ray limits on the primordial black hole abundance from Hawking radiation," Phys.Lett.B 808 (2020) 135624, arXiv:1906.10113v2 [astro-ph.CO].
\bibitem{dark}A. Coogan, L. Morrison and S. Profumo, "Direct Detection of Hawking Radiation from Asteroid-Mass Primordial Black Holes," Phys. Rev. Lett. 126, 171101 (2021), arXiv:2010.04797v1 [astro-ph.CO].



\bibitem{hayden}P. Hayden and J. Preskill, "Black holes as mirrors: quantum information in random subsystems," JHEP 0709:120,2007, arXiv:0708.4025v2 [hep-th].
\bibitem{scram}N. Lashkari, D. Stanford, M. Hastings, T. Osborne and P. Hayden, "Towards the fast scrambling conjecture," JHEP 2013:22, 2013, arXiv:1111.6580v2 [hep-th].
\bibitem{water}S. Weinfurtner, E. W. Tedford, M. C. J. Penrice, W. G. Unruh and G. A. Lawrence, "Measurement of stimulated Hawking emission in an analogue system," Phys.Rev.Lett.106:021302,2011, arXiv:1008.1911v2 [hep-th].
\bibitem{cloak}C. García-Meca, S. Carloni, C. Barceló, G. Jannes, J. Sánchez-Dehesa and A. Martínez ,''Analogue Transformations in Physics and their Application to Acoustics," Scientific Reports volume 3, Article number: 2009 (2013).
\bibitem{a}C. Barcelo, S. Liberati, M. Visser, "Analogue Gravity," 	arXiv:0505065 [gr-qc].
\bibitem{exp}M. J. Jacquet, S. Weinfurtner and F. König, "The next generation of analogue gravity experiments," Phil.Trans.Roy.Soc.Lond.A 378 (2020),  arXiv:2005.04027v2 [gr-qc].

\bibitem{infoana}S. Parvizi and M. Shahbazi, "Analogue gravity and the island prescription," Eur. Phys. J. C (2023) 83: 705, arXiv:2302.08742 [hep-th].


\bibitem{sc}S. A. Hartnoll, C. P. Herzog and G. T. Horowitz, "Holographic Superconductors," JHEP 0812:015,2008, 	arXiv:0810.1563 [hep-th].
\bibitem{sc1}S. A. Hartnoll, A. Lucas and S. Sachdev, "Holographic quantum matter,"  arXiv:1612.07324 [hep-th].

\bibitem{conformalana}R. G. Leigh, A. C. Petkou and P. M. Petropoulos, "Holographic Fluids with Vorticity and Analogue Gravity," JHEP 11 (2012) 121 , arXiv:1205.6140 [hep-th].
\bibitem{dumb}S. R. Das, A. Ghosh, J.-H. Oh and A. D. Shapere, "On Dumb Holes and their Gravity Duals," JHEP 04 (2011) 030, arXiv:1011.3822 [hep-th].




\bibitem{des}C. Y. Lin, D. S. Lee and R. J. Rivers, "The role of Causality in Tunable Fermi Gas Condensates," J. Phys. : Condens. Matter 25, 404211 (2013), arXiv:1205.0133 [gr-qc].
\bibitem{sch}C. Barcelo, S. Liberati and M. Visser, "Analog gravity from Bose-Einstein condensates," Class. Quant. Grav. 18, 1137 (2001), arXiv:0011026 [gr-qc].
\bibitem{bb}S. Hossenfelder, "Analog Systems for Gravity Duals," arXiv:1412.4220v2 [gr-qc].
\bibitem{ac}S. Hossenfelder and T. Zingg, "Analogue Gravity Models From Conformal
Rescaling," arXiv:1703.04462v2 [gr-qc].
\bibitem{particle} M. Visser and C. Molina-Paris, "Acoustic geometry for general relativistic barotropic irrotational fluid flow," 	New J.Phys.12:095014,2010,  arXiv:1001.1310v2 [gr-qc].
\bibitem{ra}X.-H. Ge, J.-R. Sun, Y. Tian, X.-N. Wu and Y.-L. Zhang, "Holographic Interpretation of Acoustic Black Holes," arXiv:1508.01735v2 [hep-th].

\bibitem{degree}S.-J. Zhang, "Complexity and phase transitions in a holographic QCD model," Nucl.Phys.
B929 (2018) 243-253, arXiv:1712.07583 [hep-th].

\bibitem{fcs}A. Akhavan, M. Alishahiha, A. Naseh and H. Zolfi, "Complexity and Behind the Horizon Cut Off," JHEP12(2018)090, arXiv:1810.12015 [hep-th].


\bibitem{lya}A. Bhattacharyya, S. Das, S. S. Haque and B. Underwood, "The Rise of Cosmological Complexity: Saturation of Growth and Chaos," Phys. Rev. Research 2, 033273 (2020),  arXiv:2005.10854v1 [hep-th].

\bibitem{cosmology}A. Bhattacharyya, S. Das, S. S. Haque and B. Underwood, "Cosmological Complexity," Phys. Rev. D 101, 106020 (2020), arXiv:2001.08664v2 [hep-th].


\bibitem{prob}E. Jørstad, R. C. Myers and S.-M. Ruan, "Complexity=Anything: Singularity Probes," JHEP 07 (2023) 223, arXiv: 2304.05453 [hep-th].


\bibitem{v}H.-S. Liu, H. Lü, L. Ma and W.-D Tan, "Holographic Complexity Bounds," JHEP 07 (2020) 090, arXiv:1910.10723 [hep-th].

\bibitem{unr}U. W.G., "Experimental black hole evaporation," Phys. Rev. Lett. 46, 1351-1353 (1981).
\bibitem{son} P. Kovtun, D. T. Son and A. O. Starinets, "Viscosity in strongly interacting quantum field theories from black hole physics," Phys.Rev.Lett. 94 (2005) 111601, arXiv:0405231 [hep-th].
\bibitem{natsume}M. Natsuume, "AdS/CFT Duality User Guide," Lect.Notes Phys. 903 (2015) pp.1-294, arXiv:1409.3575v4 [hep-th].

\bibitem{viscos}M. Visser, "Acoustic black holes: horizons, ergospheres, and Hawking radiation," Class.Quant.Grav. 15 (1998) 1767-1791, arXiv:9712010 [gr-qc].


\bibitem{vio1}A. Dobado, F. J. Llanes-Estrada and J. M. T. Rincon, "The Status of the KSS bound and its possible violations: How perfect can a fluid be?," AIP Conf.Proc. 1031 (2008) 1, 221-231,  arXiv:0804.2601v1 [hep-ph].
\bibitem{vio2}A. Dobado and F. J. Llanes-Estrada, "On the violation of the holographic viscosity versus entropy KSS bound in non relativistic systems," Eur.Phys.J.C 51 (2007) 913-918, arXiv:0703132v3 [hep-th].
\bibitem{vio3}L. Albertea, M. Baggiolib and O. Pujolàsb, "Viscosity bound violation in holographic solids and the viscoelastic response," JHEP07(2016)074, arXiv:1601.03384v4 [hep-th].
\bibitem{poly}J. Christians, "Approach for Teaching Polytropic Processes Based on the Energy Transfer Ratio," International Journal of Mechanical Engineering Education, 40(1), 53–65.
\bibitem{eos1}V.V. Begun, M.I. Gorenstein and O.A. Mogilevsky, "Modified Bag Models for the Quark Gluon Plasma Equation of State," Int.J.Mod.Phys.E20:1805-1815,2011,  arXiv:1004.0953v3 [hep-ph].
\bibitem{eos2}S. M. Sanches Jr., F. S. Navarra and D. A. Fogaça, "The quark gluon plasma equation of state and the expansion of the early Universe," Nuclear Physics A 937 (2015), 1, arXiv:1410.3893v2 [hep-ph].
\bibitem{gbv}M. Brigante, H. Liu, R. C. Myers, S. Shenker and S. Yaida, "Viscosity Bound Violation in Higher Derivative Gravity," 	Phys.Rev.D77:126006,2008,  arXiv:0712.0805v3 [hep-th].

\bibitem{gbv2}R.-G. Cai, Z.-Y. Nie, N. Ohta and Y.-W. Sun, "Shear Viscosity from Gauss-Bonnet Gravity with a
Dilaton Coupling," Phys.Rev.D79:066004,2009,  arXiv:0901.1421v2 [hep-th].

\bibitem{gbv3}A. Buchel, R. C. Myers and A. Sinha, "Beyond eta/s = 1/4pi," JHEP 0903:084,2009,  arXiv:0812.2521v3  [hep-th].

\bibitem{gbv4}R.-G. Cai, Z.-Y. Nie and Y.-W. Sun, "Shear Viscosity from Effective Couplings of Gravitons," 	Phys.Rev.D78:126007,2008, arXiv:0811.1665v2 [hep-th].

\bibitem{fg}M. Rangamani, "Gravity $\&$ Hydrodynamics: Lectures on the fluid-gravity correspondence,"  arXiv:0905.4352v3 [hep-th].


\bibitem{general}N. Pinzani-Fokeeva and M. Taylor, "Towards a general fluid/gravity correspondence," Phys. Rev. D 91, 044001 (2015),  arXiv:1401.5975v2 [hep-th].

\bibitem{kss-1}G. Koutsoumbas, E. Papantonopoulos and G. Siopsis, "Shear Viscosity and Chern-Simons Diffusion Rate from Hyperbolic Horizons," Phys.Lett.B 677 (2009) 74-78, arXiv:0809.3388v2 [hep-th].
\bibitem{gbboundary}M. Brigante, H. Liu, R. C. Myers, S. Shenker and S. Yaida, "Viscosity Bound Violation in Higher Derivative Gravity,"  Phys.Rev.D77:126006,2008, arXiv:0712.0805v3 [hep-th].
\bibitem{etgb}D. Glavan and C. Lin, "Einstein-Gauss-Bonnet gravity in 4-dimensional space-time," Phys. Rev. Lett. 124, 081301 (2020),  arXiv:1905.03601v3 [hep-th].






\end{thebibliography}
\end{document}